\def \vamp { v_{\rm amp}}
\def \vsurf { v_{\rm surf}}
\def \Bbar {{\overline {\bm B}}}
\def \Abar {{\overline {\bm A}}}
\def \Jbar {{\overline {\bm J}}}

\def \Beq {B_{\rm eq}}

\def \rzero {r_0}

\def \Ubar {{\overline {\bm U}}}
\def \Uwind {{\bm U}^{\rm wind}}
\def \Ucirc {{\bm U}^{\rm circ}}
\def \Ushear {{\bm U}^{\rm shear}}

\def \curl {{\bm \nabla} \times}

\def \lap {\nabla^2}

\newcommand{\bra}[1]{\langle #1\rangle}
{}

\def \Rm  {\mbox{Re}_{\rm M}}

\def \kf  {k_{\rm f}}
\def \curl {{\bm \nabla}\times}

\def \etat {\eta_{\rm t}}

\def \Rm  {\mbox{Re}_{\rm M}}

\def \kf  {k_{\rm f}}

\def \alpk {\alpha_{\rm K}}
\def \alpm {\alpha_{\rm M}}

\newcommand{\pd}{\partial}

\newcommand{\apj}{ApJ}

\documentclass{aa}
\usepackage{bm}
\usepackage{showlabels}[1]
\usepackage{lscape,rotating}
\usepackage{natbib}
\usepackage{hyperref}
\usepackage{amsmath,amssymb}
\usepackage{times}
\usepackage{graphicx}
\usepackage{fancyhdr}
\usepackage{url}
\graphicspath{{./plots/}{./png/}}
\def \Rm  {\mbox{Re}_{\rm M}}

\def \kf  {k_{\rm f}}

\def \etat{\eta_{\rm t}}

\newcommand{\meanEMF}{\overline{\mbox{\boldmath ${\cal E}$}}{}}{}
\title{Alleviating $\alpha$ quenching by solar wind and meridional flows}
\authorrunning{D. Mitra et al.}
\author{Dhrubaditya Mitra\inst{1}, David Moss\inst{2},
Reza Tavakol\inst{3}, and Axel Brandenburg\inst{1,4} }
\institute{NORDITA, AlbaNova University Center, Roslagstullsbacken 23, SE-10691 Stockholm, Sweden
\and School of Mathematics, University of Manchester, Oxford Road,
       Manchester M13 9PL, UK
\and 
Astronomy Unit, School of Mathematical Sciences, Queen Mary University of London,
Mile End Road, London E1 4NS, UK
\and
Department of Astronomy, Stockholm University, SE 10691 Stockholm, Sweden
}
\date{\today,~ $ $Revision: 1.90 $ $}
\begin{document}
\abstract
{}
{We study the ability of magnetic helicity expulsion
to alleviate catastrophic $\alpha$-quenching in mean field dynamos
in two--dimensional spherical wedge domains.}
{Motivated by the physical state of the outer regions of the Sun,
we consider $\alpha^2\Omega$
mean field models with a dynamical $\alpha$ quenching.
We include two mechanisms which have the potential to
facilitate helicity expulsion, namely advection by a 
mean flow (``solar wind'') and meridional circulation.}
{We find that a wind alone can prevent catastrophic quenching, with the field
saturating at finite amplitude. In certain parameter ranges, the
presence of a large-scale meridional circulation can reinforce this
alleviation. However, the saturated field strengths are typically below 
the equipartition field strength. We discuss possible mechanisms that 
might increase the saturated field.}
{}
%%%%%%%%%%%%%%
\maketitle
%----------------------
\section{Introduction}
%----------------------
Mean field dynamo models have provided an important framework
for studying the generation of large-scale  astrophysical
magnetic fields and their spatio-temporal dynamics.    
However, 
 these widely used models have been presented with a serious 
challenge -- namely
the so called catastrophic $\alpha$ quenching \citep{GD94}.
In the mean field (MF)  context this effect, which is a consequence of the
conservation of magnetic helicity \citep{KR80,ZRS83},
manifests  itself as the decrease of the $\alpha$--effect with
increasing magnetic Reynolds number $\Rm$ \citep{VC92,CH96}
at finite field strength.
In models without
magnetic helicity fluxes, the quenching of $\alpha$
can become severe, with $\alpha$ decreasing as $\Rm^{-1}$  -- truly 
catastrophic for dynamo action
in the Sun, stars and galaxies where
the Reynolds numbers are all very large ($>10^9$).
This catastrophic quenching is captured by mean-field
models which use dynamical alpha quenching, such as
that considered by \cite{BB02}. 
This catastrophic quenching is independent of the details of
the dynamo mechanism and is a direct effect of the conservation
of magnetic helicity, see, e.g., \cite{BK07} who have demonstrated
catastrophic quenching for non-local alpha effect or 
\cite{CGB2010} who have demonstrated the occurrence of catastrophic 
quenching in distributed dynamos. 
It has been suggested that the quenching may be alleviated by
the expulsion of magnetic helicity through open 
boundaries \citep{BF00,KMRS00}.
At least three different physical mechanisms may help
in the expulsion of small scale magnetic helicity:
(a) large scale shear \citep{VC01,SB04,BS04,MS2010};
(b) turbulent diffusion of magnetic helicity \citep{MCCTB2010};
(c) non-zero mean flow out from a boundary of the domain, e.g. a wind. 
%A number of studies have demonstrated the possibility of this
%AB: David's suggestion
A number of recent studies have demonstrated the possibility of this
alleviation of quenching for solar \citep{CGB2010,CBG2010,GCB2010} and 
galactic dynamos \cite[e.g.][]{SSSB06}.

In this paper we study the effects of a number of mechanisms
which may facilitate the expulsion 
of magnetic helicity from the dynamo region. Initially we consider
the effects of advection by a mean flow in a similar manner to \cite{SSSB06};
see also the recent study in a one dimensional model by \cite{BCC09}.
We envisage that in the Sun the wind could be loaded with magnetic
helicity through coronal mass ejections \citep{BB03}.
Another potentially important mechanism 
is meridional circulation. The presence of
such a circulation in the Sun is supported by
a number of  observations which have found
evidence for a near-surface poleward flow
of $10 - 20$~ms$^{-1}$. Even though the
corresponding compensating equatorward flow
has not yet been detected,
it is however assumed
it must exist because of mass conservation. 
Substantial effort has recently gone into
the construction of flux transport dynamo models which
differ from the usual $\alpha\Omega$ dynamos
by including an additional advective transport
of magnetic flux by meridional circulation.
\citep[see e.g.][for a recent summary]{Dikpati-Gilman-09}.
If magnetic flux is advected by meridional circulation,
it can be expected that
such a circulation will also
transport magnetic helicity to the surface layers,
which might thus facilitate its subsequent expulsion by the wind.
We therefore study the effects of meridional circulation
on the quenching. 

The structure of the paper is as follows.
In Section~\ref{Model} we introduce  our model
and its various ingredients.
Section~\ref{Results} contains our results, and we give a short summary
here. First we consider our model with an imposed
wind but no meridional circulation. We show that a strong enough
wind that penetrates deeply enough into the convection zone can indeed
alleviate quenching. We then make a systematic study of the
alleviation of quenching as a function of the two parameters
specifying the wind, namely the maximum velocity and the depth down to which
the wind penetrates the convection zone. 
Next we select a particular set of these two parameters such
that for large $\Rm$ there is no alleviation of quenching.
We then introduce a meridional circulation and show that
a combination of a wind and circulation is able to
limit the quenching in cases where the wind alone
cannot. We further study the effect of the characteristic
velocity of meridional circulation on quenching. 
Our conclusions are presented in Section~\ref{Conclusions}.

%-------------------
\section{The model}
\label{Model}
%-------------------

We study two--dimensional (axisymmetric) mean field models in a 
spherical wedge domain,
$r_1\le r\le r_2$, $\theta_1\le\theta \le \pi/2$, 
where $r,\theta,\phi$ are spherical polar coordinates.
The choice of this  ``wedge" shaped domain is motivated by recent
Direct Numerical Simulations (DNSs) of forced and convective dynamos in 
spherical wedges cut from spheres \citep{MTKB2010,KKBMT2010}, and the
intention to make a similar development of this work.
  
We consider an $\alpha^2 \Omega$ mean field model 
with a ``dynamical alpha'' in the presence of an
additional mean flow $\Ubar$. In the simplest case,
where we consider no wind and no meridional circulation,
the mean flow is  in the form
of a uniform rotational shear given by 
$ \Ubar = \Ushear = \bm{\hat\phi}S ( r - \rzero)\sin\theta$.
For the more realistic cases we use
\begin{equation}
\Ubar = \Ushear+\Uwind+\Ucirc ,
\label{eq:ubar}
\end{equation}
where $\Uwind$ and $\Ucirc$ are respectively the large-scale
velocity of the wind and circulation. The particular forms
we use are given in Sections~\ref{sec:wind} and 
\ref{sec:circulation} below. Thus, we integrate
\begin{eqnarray}
\partial_t \Bbar &=& \curl (\Ubar\times\Bbar+\meanEMF) + \eta\lap \Bbar, 
\label{mfeqn}\\
\partial_t \alpha_{\rm M}&=& -2\eta\kf^2 \left( 
                                    \frac{\meanEMF\cdot\Bbar}{\Beq^2} +
                                    \frac{\alpha_{\rm M}}{\Rm} \right)
-\vec{\nabla}\cdot(\alpha_{\rm M}\Ubar) ,
\label{alphaeqn}
\end{eqnarray}
where $\meanEMF=\alpha \Bbar -\etat \Jbar$, and
$\alpha = \alpha_{\rm M} + \alpha_{\rm K}$
is the sum of the  magnetic and kinetic $\alpha$--effects respectively.
The magnetic Reynolds number,
$\Rm/3\equiv\etat/\eta$ and $\Beq$ is the equipartition field strength.
We take $\etat=1$, $\Beq=1$ and $\kf=100$ in our simulations.
 Here Eq.~(\ref{mfeqn}) is the standard induction equation for
mean field models and Eq.~(\ref{alphaeqn}) describes the dynamical
evolution of $\alpha$; see \cite{BB02}. 
The last term in the right hand side of Eq.~(\ref{alphaeqn})
models the advective flux of magnetic helicity.

%We solve Equations (\ref{mfeqn}) and (\ref{alphaeqn}) by 
%AB: changed
We solve Equations (\ref{mfeqn}) and (\ref{alphaeqn}) using
the {\sc Pencil Code}%
\footnote{\texttt{http://pencil-code.googlecode.com/}} 
%which uses a sixth order centered finite-difference
which employs a sixth order centered finite-difference
method to evaluate the spatial derivatives and a third
order Runge-Kutta scheme for time evolution. 

Our aim here is to study the effects of the various mechanisms discussed
above in alleviating the catastrophic quenching of the
magnetic field as $\Rm$  increases.
%-----------------------------------------
\subsection{The wind and the ``corona''}
\label{sec:wind}
%-----------------------------------------
In order to include the effects of the solar wind we must include 
an outer region in our model through which the wind flows,
by extending the outer boundary beyond the convection zone to
radius $r_3 > r_2$.  We shall refer to the region $ r_2 \le r\le r_3$
as the `corona'.
We take the wind to be strong in the corona and to grow weaker as we go
into the convection zone.
 This is represented by choosing the following form
for $\Uwind$, 
\begin{eqnarray}
\Uwind_r &=&  \frac{1}{2} U_0 \left [ 1+\tanh\left(\frac{r-r_2}{w} \right) \right] , \label{uwindr}\\
 \Uwind_\theta &=& 0, \label{uwindth}\\
\Uwind_\phi &=& 0,
\end{eqnarray}
where $U_0$ and $w$ are control parameters which determine the
strength of the wind speed and its depth of penetration into the convection zone
respectively. Larger values of $w$ correspond to deeper penetration.
We let the kinematic $\alpha$--effect to
go to zero in the corona by choosing 
\begin{equation}
\alpk = -\frac{\alpha_0}{2}\tanh \left( \frac{\theta-\pi/2}{0.05} \right)
             \left[1 -\tanh\left(\frac{r-r_2}{w_{\alpha}} \right) \right] ,
\label{ystep_xcutoff}
\end{equation}
with $\alpha_0 =16$.
%-----------------------------------------
\subsection{The meridional circulation}
\label{sec:circulation}
\begin{figure}
\includegraphics[width=0.8\linewidth]{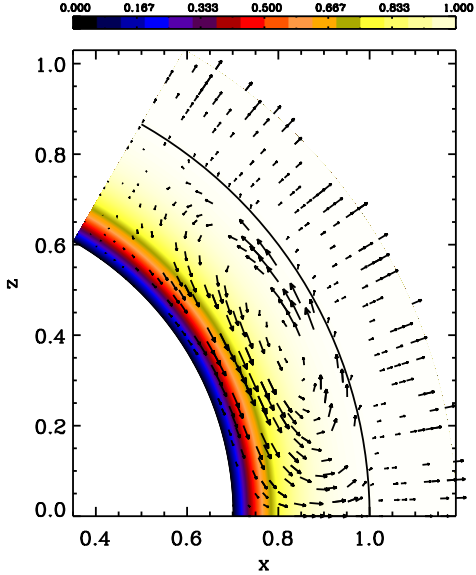}
\caption{Plot of the velocity field: the arrows show the meridional
circulation and the wind, and the contours show the angular velocity.
The solar radius is taken to be unity. Although our domain extends out to
$5$ solar radii, for clarity only a part of it is shown here.
The curve at unit radius denotes the surface of the Sun. }

\label{fig:usnap}
\end{figure}

\begin{figure}
\includegraphics[width=0.8\linewidth]{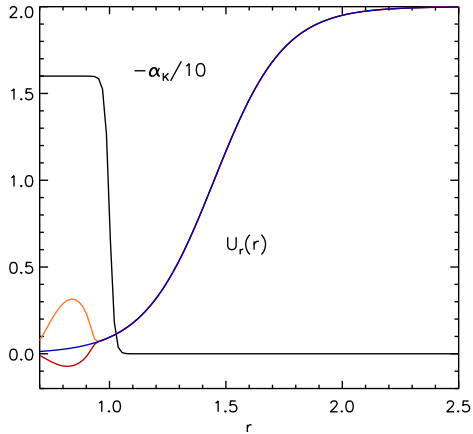}
\caption{The kinetic alpha effect, $\alpk$, and wind radial
%velocity, $U_r$ as a function of radial coordinate $r$ for three
%AB: Reza's change
velocity, $U_r$, as a function of radial coordinate $r$ for three
different latitudes, equator (upper curve), mid-latitude (middle curve) and
latitude of upper boundary (lower curve). 
Note that the curves for the radial velocities differ only in $r <   1$,
where the meridional circulation is non-zero. }
\label{fig:profile}
\end{figure}

%-----------------------------------------
We consider the effects of a
meridional circulation, by including a velocity $\Ucirc$ given by
\begin{eqnarray}
\Ucirc_r&=&\vamp g(r)\frac{1}{\sin\theta}\frac{\partial}
{\partial\theta}\left(\sin\theta\,\psi\right), \\
\Ucirc_\theta&=&-\vamp g(r)\frac{1}{r}\frac{\partial}{\partial r}(r\psi),  \\
\psi&=&\frac{f(r)}{r}\sin^2(\theta-\theta_1)\cot\theta, \\
f(r)&=& (r-r_2)(r-r_1)^2, \\
g(r)&=& \frac{1}{2}\left[1 -\tanh\left(\frac{r-r_2}{w_{\rm circ}} \right) \right], \\
\Ucirc_{\phi} &=&  0. 
\label{circulation}
\end{eqnarray}
Here $\vamp$ is a parameter controlling the magnitude of circulation
speed and $w_{\rm circ}$ determines the effective depth of penetration
of the
circulation into $r> r_2$.
As a characteristic speed of circulation, $v_{\rm circ}$, we take the
maximum absolute magnitude of the $\theta$ component of
$\Ucirc$ at $r=r_2$, i.e. at the surface of
the Sun. Helioseismology shows this velocity to be about $10$ to $20$
metres per second in the Sun.
A typical velocity field is shown in Fig.~\ref{fig:usnap},
and the profile of $\alpk$ is shown in Fig.~\ref{fig:profile}.
In these Figures, the  parameters are chosen to be
$\alpha_0=16$, $w_{\alpha}=0.2$, $U_0=2$, $r_2=1.5$, $w=0.3$,
$\vamp=75$, $r_{\rm circ}=0.98$, $w_{\rm circ}=0.02$.

%%%%%%%%%%%%%%%%%%%%%%%%%%%%%%%%%%%%%%%%
%----------------------------
%%%%%%%%%%%%%%%%%%%%%%%%%%%%%%%%%%%%%%%%
%------------------------------ 
\subsection{Boundary conditions}
\label{sec:boundary}
%------------------------------
For the magnetic field we use perfect conductor boundary conditions both at
the base of the convection zone (at $r=r_1$) and at the lateral 
boundary at the higher latitude ($\theta = \theta_1$).
We assume the magnetic field to be antisymmetric about
the equator ($\theta = \pi/2$),
and at the outer radial boundary of the corona ($r=r_3$) we use the 
normal field condition.
In terms of the magnetic vector potential 
$\Abar = (A_r, A_\theta, A_{\phi})$, where
$\Bbar = \curl \Abar$, these conditions become  
\begin{eqnarray}
&&\frac{\pd A_r}{\pd r}= A_\theta=A_\phi=0 \quad (r=r_1), \label{bc1r1}\\
&&A_r=0, \quad \frac{\pd A_{\theta}}{\pd r}=-\frac{A_{\theta}}{r}, \quad \frac{\pd A_{\phi}}{\pd r}=-\frac{A_{\phi}}{r} \quad (r=r_3), \label{bc1r2}\\
&&A_r=\frac{\pd A_\theta}{\pd\theta}=A_\phi=0 \quad (\theta=\theta_1), \label{bc1t1}\\
&&\frac{\pd A_r}{\pd \theta}= A_\theta = \frac{\pd}{\pd\theta}(\sin\theta A_\phi)=0 \quad (\theta=\pi/2).
\label{bc1t2}
\end{eqnarray}
For $\alpm$, 
on those boundaries where the boundary condition on the magnetic
field is ``perfect-conductor'' (i.e.\ at the bottom of the 
convection zone and at the higher latitude), we choose 
\begin{equation}
\alpm=0 .
\end{equation}
At the other two boundaries, we recall that since the PDE being solved
is of first order in space we only need to specify one condition,
which we have already imposed at the lower boundary. 
To calculate the derivative
at the outer boundary we therefore just extrapolate the solution from
inside to outside by a second order polynomial
extrapolation. 
This is equivalent to using second order one sided
finite difference at these boundaries.

As the initial condition for the magnetic field we choose our 
seed magnetic vector potential from a random Gaussian distribution 
with no spatial correlation and root-mean-square
value of the order of $ 10^{-4}$ times the equipartition field strength.
Also, initially we take $\alpm = \alpha - \alpk = 0$.

%------------------
\section{Results}
\label{Results}
\begin{figure}
\includegraphics[width=0.9\linewidth]{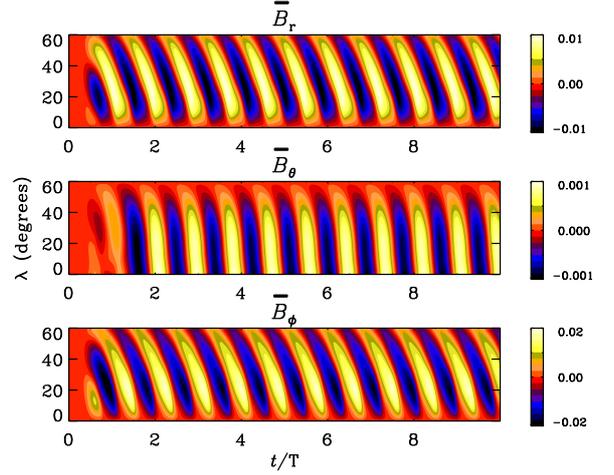} 
\caption{Spacetime diagrams for the three components of the magnetic field.
These plots are for the set of parameters corresponding to the velocity
%field and $\alpk$, shown in Figs.~\ref{fig:usnap} and ~\ref{fig:profile}.}
%AB: David's change
field and $\alpk$ shown in Figs.~\ref{fig:usnap} and ~\ref{fig:profile}.}
\label{fig:bfly_typical}
\end{figure}

%------------------
In order to demonstrate that our dynamo
is excited, and displays both oscillations and equatorward migration, we first
use the velocity field and the kinetic $\alpha$ profile
shown in Fig.~\ref{fig:usnap},
with $\Rm=3\times10^{2}$
and solve Eqs.~(\ref{mfeqn}) and (\ref{alphaeqn}) simultaneously. The resulting
space-time diagram for the three components of the magnetic field is shown
in Fig.~\ref{fig:bfly_typical}. This is a typical example of the ``butterfly''
diagrams that are obtained with this model.
%----------------------------------
%--------------------------------------

\begin{figure}
\includegraphics[width=0.9\linewidth]{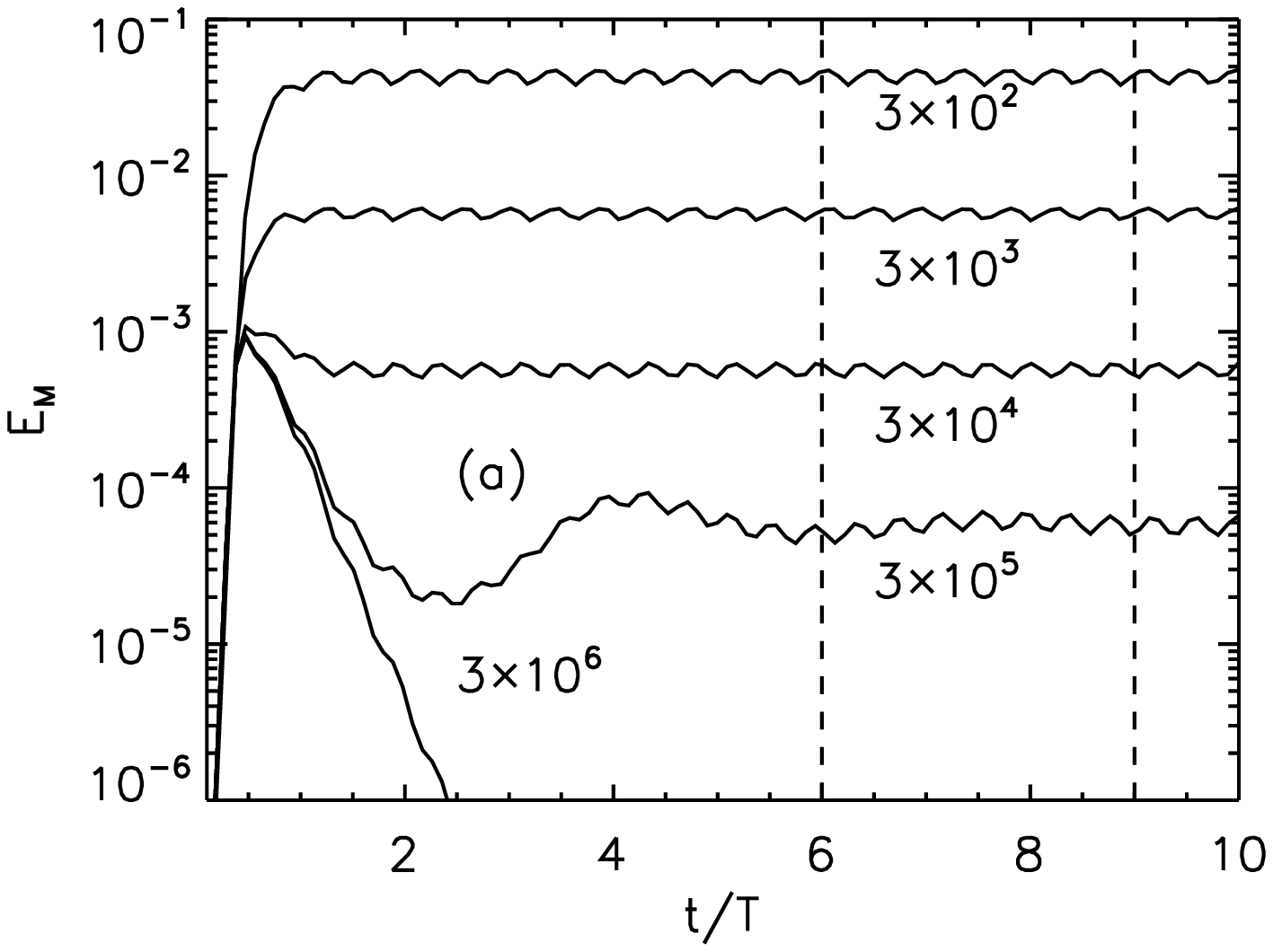} 
\includegraphics[width=0.9\linewidth]{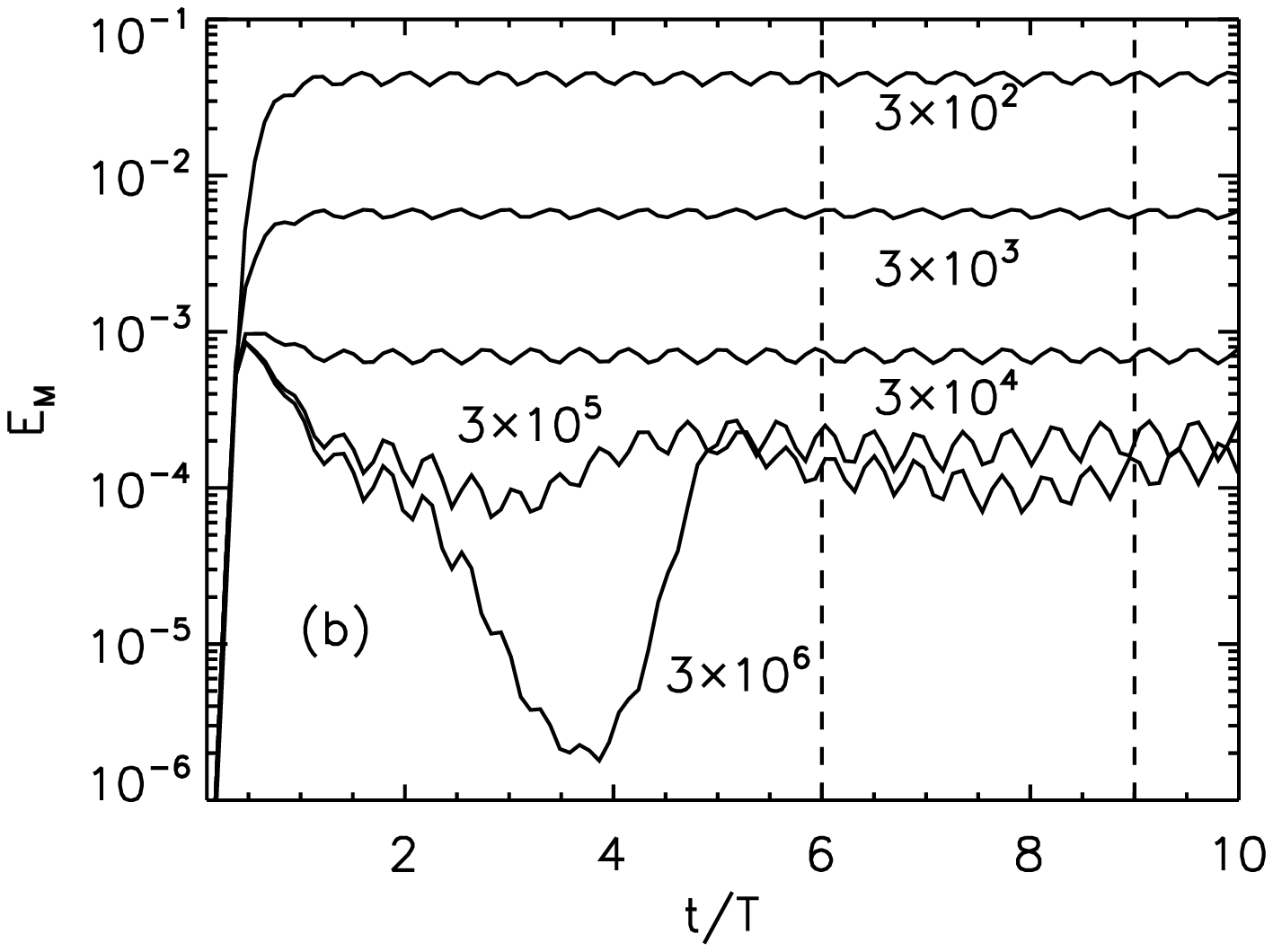} 
\caption{(a) Time series of $E_M$ 
for $U_0=0$ and $\vamp=0$, i.e.\ no wind or circulation, 
for 5 different values of $\Rm$.
(b) The corresponding plot 
with $U_0=1$, $w=0.3$ and $\vamp=75$. The time-averaged magnetic energy
$\langle E_M \rangle_t$ is calculated by time-averaging these time-series
between the two dashed vertical lines. The other parameters used are
$\alpha_0=16$, $w_{\alpha}=0.2$, $r_2=1.5$, and $w=0.3$.}
\label{fig:EMvst_quench}
\end{figure}

As mentioned above, an important feature of MF dynamos in the absence 
of wind and meridional circulation (i.e. 
when $U_0=0$ and $\vamp=0$), is that they are severely quenched
as $\Rm$ increases. To show this we have plotted in 
Fig.~\ref{fig:EMvst_quench} (a) the time-series of the total magnetic energy
$ E_M  = {1\over2}\bra{\Bbar^2}$ for several values of $\Rm$. 
Here, $\bra{...}$ denotes averaging over the domain $r_1 \le r \le r_2$.
Clearly the total magnetic energy
decreases with $\Rm$. Similar quenching, as a result of  the dynamical
evolution of the alpha term,
has been seen in many different models
of the solar dynamo \citep[see, e.g.,][for some recent examples]{CGB2010,CBG2010,GCB2010},
and also in models of galactic dynamos \citep{SSSB06}.

%----------------------------------
\begin{figure}
\includegraphics[width=0.9\linewidth]{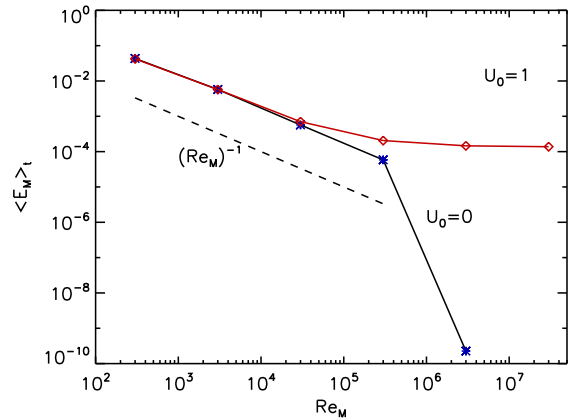} 
\caption{The time-averaged magnetic energy as a function of $\Rm$ for 
$U_0 =0$ (no wind) and $U_0 =1$ and $w=0.3$ and $\vamp=75$. }
\label{fig:noquench}
\end{figure}

%--------------------------------------
To substantiate this further we plot in Fig.~\ref{fig:noquench}
the time-averaged magnetic energy $\langle E_M \rangle_t$ as a function of 
$\Rm$, where the time averaging is done over several diffusion times ($T$) 
in the saturated nonlinear stage 
(i.e.\ after the kinematic growth phase is over). 
Time averaging is here indicated by the subscript $t$ after the averaging sign.
As can be seen in the absence of wind, i.e. with $U_0 =0$,
such time-averaged energy falls off approximately as $\Rm^{-1}$. 
This gives a quantitative measure of the quenching. 
(The point at $\Rm=2\times 10^6$ appears anomalous; we believe this is because
we have not run the code for long enough to achieve the final saturated state.)

To demonstrate the ability of the wind
and circulation to act together 
to alleviate quenching, we have 
plotted in Fig.~\ref{fig:EMvst_quench} (b) the time-series of
$E_M$ for several different values of $\Rm$,
in the presence of the wind (with $U_0 =1$, $\vamp=75$ and depth parameter $w=0.3$).
The dependence of the time-averaged magnetic energy $\langle E_M \rangle_t$ 
on $\Rm$ in this case is also plotted in Fig.~\ref{fig:noquench}.
Comparing Fig.~\ref{fig:EMvst_quench} (b) with Fig.~\ref{fig:EMvst_quench} (a)
and also comparing the two lines in~Fig.~\ref{fig:noquench}
we clearly see that with the parameters chosen the
wind in conjunction with the circulation is capable of alleviating quenching.
This is one of our principal results.
Note that the saturated mean field energy that we observe at large $\Rm$ is still
rather small, only slightly exceeding $10^{-4}$ of the equipartition value.
%----------------------------------
%--------------------------------------

\begin{figure}
\includegraphics[width=0.9\linewidth]{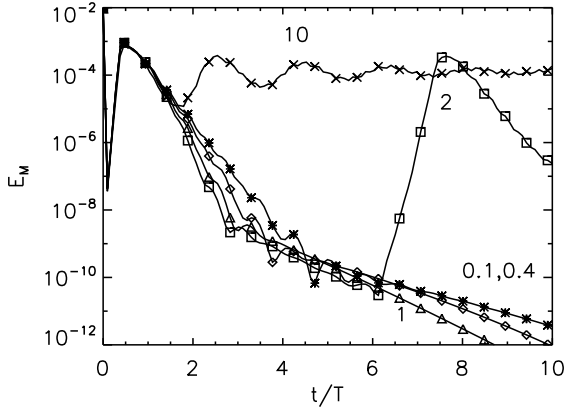} 
\caption{Time series of $E_M$ for 5 different values of $U_0$,
namely,  
$U_0=0.1 (\ast), 0.4 (\lozenge) , 1 (\vartriangle)
%     2. (\square)$  and     $10. (\times)$
%AB: remove periods
2 (\square)$ and $10 (\times)$
%with all other parameters held fixed, in particular, $\Rm=10^7$, $w=0.3$.
%AB: David's change
with all other parameters held fixed, in particular $\Rm=10^7$, $w=0.3$.
}
\label{fig:uzero_dep}
\end{figure}

Next we attempt to isolate the role of each parameter in our model. 
First we make a detailed systematic study of how
quenching depends on the two parameters 
$U_0$ and $w$ of our model,
for a fixed value of $\Rm = 10^7$ and zero circulation, $\vamp=0$.
For each pair of parameters we ran our code for up to $50$ diffusion times.
In some cases the time series of $E_M$ declines as a function
of time initially, but at larger times recovers to unquenched
values, e.g.\ $U_0=1$ in Fig.~\ref{fig:uzero_dep}. In some other
cases we observe that the recovery is merely temporary and at large times
$E_M$ goes to zero. 
As an example we first show in Fig.~\ref{fig:uzero_dep} the time-series
of $E_M$ for various values of $U_0$, for 
a fixed $w = 0.3$. 
Clearly, as the wind velocity increases the transport of magnetic
helicity out of the domain  at first  becomes more efficient and
we observe less quenching.
But this alleviation of quenching must have its limits because
for a large enough wind speed the magnetic field itself will be advected
out of the domain faster than it is generated,  thus killing the dynamo
\citep[see, e.g.,][]{SSSB06,Betal93,Metal2010}.
However with penetration factor $w=0.3$ we did not find this effect,
even when $U_0=100$, but with $w=0.5$, winds with $U_0\ge 20$ kill the dynamo.
We deduce that it is necessary to advect large-scale field from a substantial
proportion of the dynamo region for the dynamo to be killed by advection.

\begin{figure}
\includegraphics[width=0.9\linewidth]{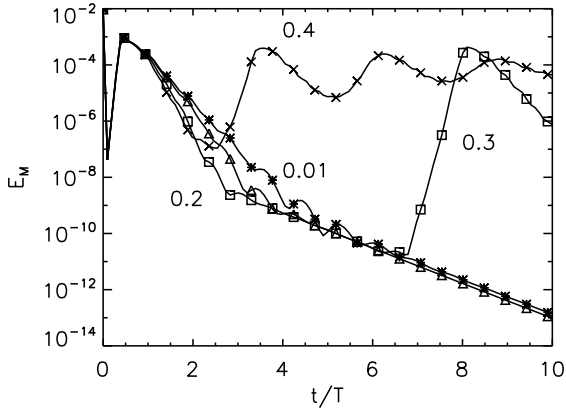}
\caption{Time series of $E_M$ for 5 different values of $w$,
[$w= 0.01 (\ast)$,$0.2 (\vartriangle)$, $0.3 (\square)$, and 
 $0.4 (\times)$.]
with all other parameters held fixed, i.e., $\Rm=10^7$, $U_0=2$
with no circulation. 
} 
\label{fig:wdep_nocirc}
\end{figure}

\begin{figure}
\includegraphics[width=0.9\linewidth]{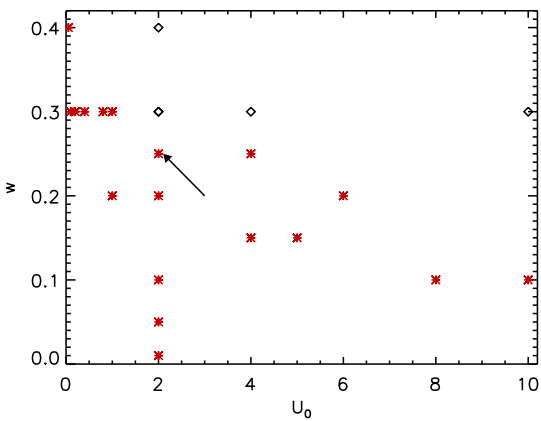}
\caption{The incidence of quenched and unquenched solutions in the
$U_0 - w $ plane. The positions of quenched models are
denoted by the symbol $\ast$,
the symbol $\lozenge$ identifies 
unquenched models. The arrow is explained in \protect Sect.~\ref{circsoln}.}
\label{fig:pdiagram}
\end{figure}

Then we consider the parameter $w$ which controls the depth of
penetration of the wind
into the convection zone. The dependence of the time-series
of magnetic energy on this parameter is shown in Fig.~\ref{fig:wdep_nocirc},
for $\Rm=10^7$ and $U_0 = 2$.
We also note that there is a subset of parameters for which
the transients are so long that it is difficult
to decide whether the asymptotic state is a
quenched dynamo or not, within reasonable integration times.
 In our parameter space, i.e. in the $U_0-w$ plane, the
positions of the quenched and unquenched runs
are shown in Fig.~\ref{fig:pdiagram}; summarizing
the dependence of quenching on these parameters.
For all the runs we label as unquenched
the butterfly diagram is also restored at large times. 
%----------------------------------
%%%%%%%%%%%%%%%%%%%%%%%%%%%%%%%%%%%%%%%%
%%%%%%%%%%%%%%%%%%%%%%%%%%%%%%%%%%%%%%%%
%%%%%%%%%%%%%%%%%%%%%%%%%%%%%%%%%%%%%%%%

\subsection{The effect of circulation}
\label{circsoln}
\begin{figure}
\includegraphics[width=0.9\linewidth]{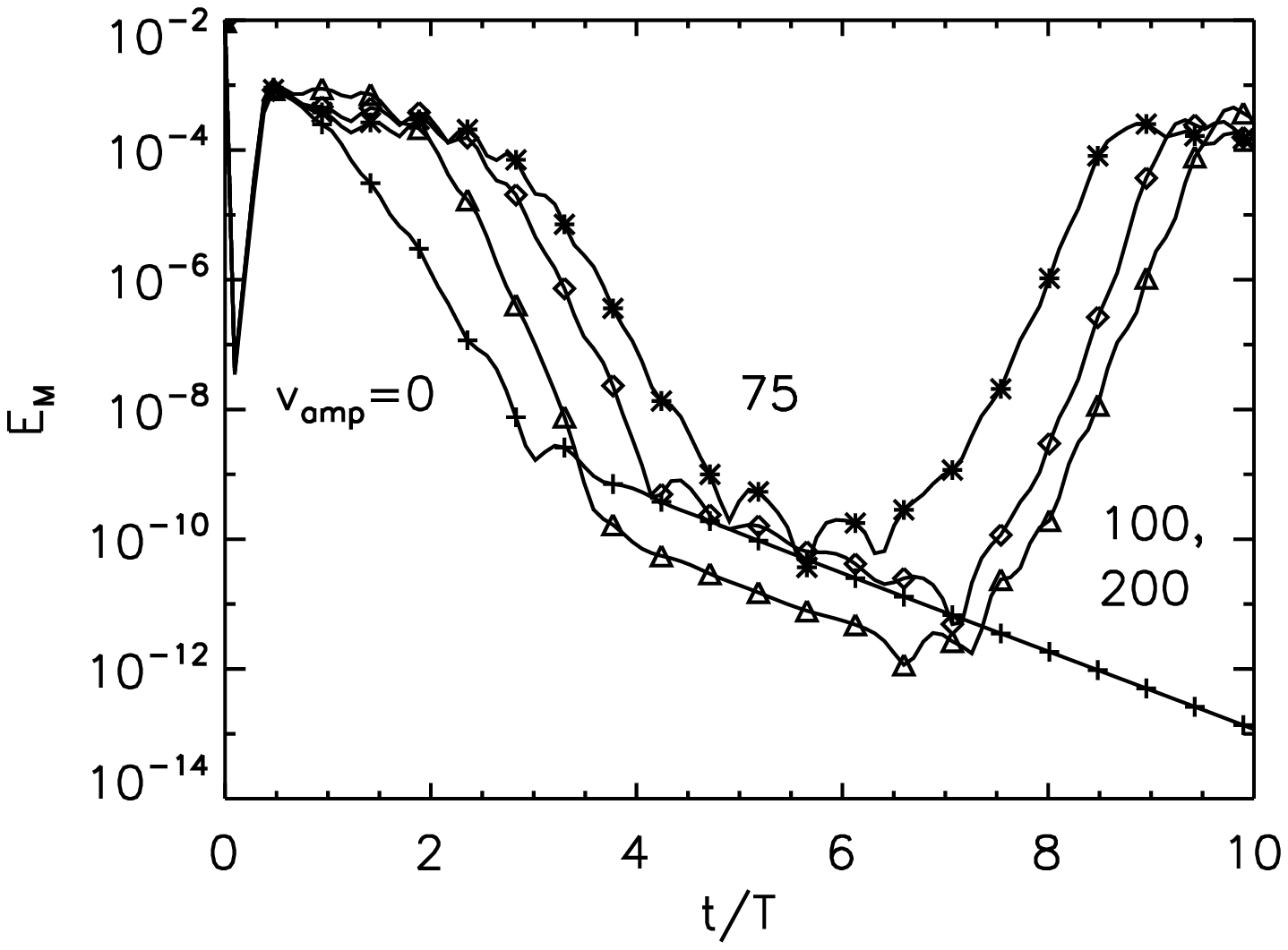}
\caption{Time series of $E_M$ as a function of time
for $U_0=2$ and $w=0.25$ with $\vamp=0 (+), 75 (\ast),
100 (\lozenge), {\rm and} 200 (\vartriangle)$. 
} 
\label{fig:circ_dep}
\end{figure}

Next we consider the effect of meridional circulation on the quenching. 
If the wind penetrates inside the convection
zone too deeply then we expect that circulation will have either no effect,
or just a marginal effect, because the wind by itself will be efficient
enough in removing small-scale magnetic helicity from deep within the domain. 
But if the wind does not penetrate so deeply, circulation may play
an important role in dredging magnetic helicity from deep inside
the domain to near the surface from where the wind can remove it.
To see whether this idea can  work,
we select one point in the phase diagram in Fig~\ref{fig:pdiagram}, 
where we obtain the quenched solution marked by the arrow.
Then we turn on the meridional circulation. 
The comparison between the time-series of $E_M$ with and 
without circulation is shown in Fig.~\ref{fig:circ_dep}.
It can be seen that the final magnetic energy reached does
not depend on the amplitude of circulation if the amplitude of
circulation is greater than a critical value. 
Note that this alleviation of quenching by the circulation only works
for those points in the $U_0-w$ parameter space  which lie close to
the boundary between the quenched and non-quenched states
in the phase diagram. For points with very 
small $w$, i.e. in cases where the wind penetrates very little into the convection zone, 
even a very strong
circulation cannot remove the quenching.
%----------------------
%----------------------

Another possible mechanism that can transport
magnetic helicity from the bulk of the convection zone to
its surface is the diffusion of magnetic helicity. This can
be described by adding the term $\kappa_{\rm t} \nabla^2 \alpm$
to the right hand side of Eq.~(\ref{alphaeqn}), where $\kappa_{\rm t}$
is an effective turbulent diffusivity of the magnetic helicity. 
Numerical simulations have estimated 
$\kappa_{\rm t} \sim 0.3\etat$ \citep{MCCTB2010}. We have checked
that such a diffusive flux of magnetic helicity can alleviate quenching
at least as effectively as the meridional circulation, in the presence
of the wind.

\begin{figure}
\includegraphics[width=0.8\linewidth]{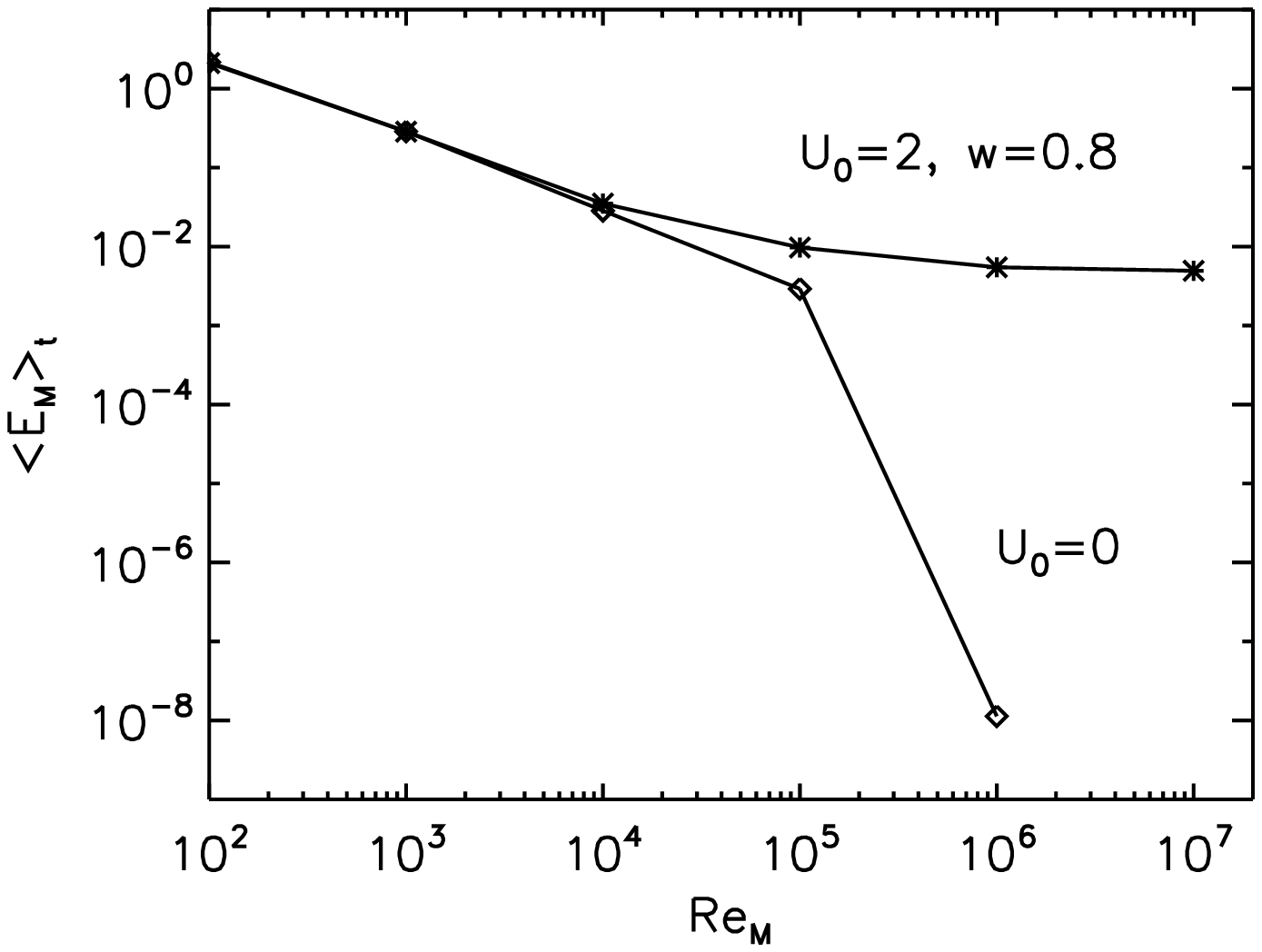}
\includegraphics[width=0.8\linewidth]{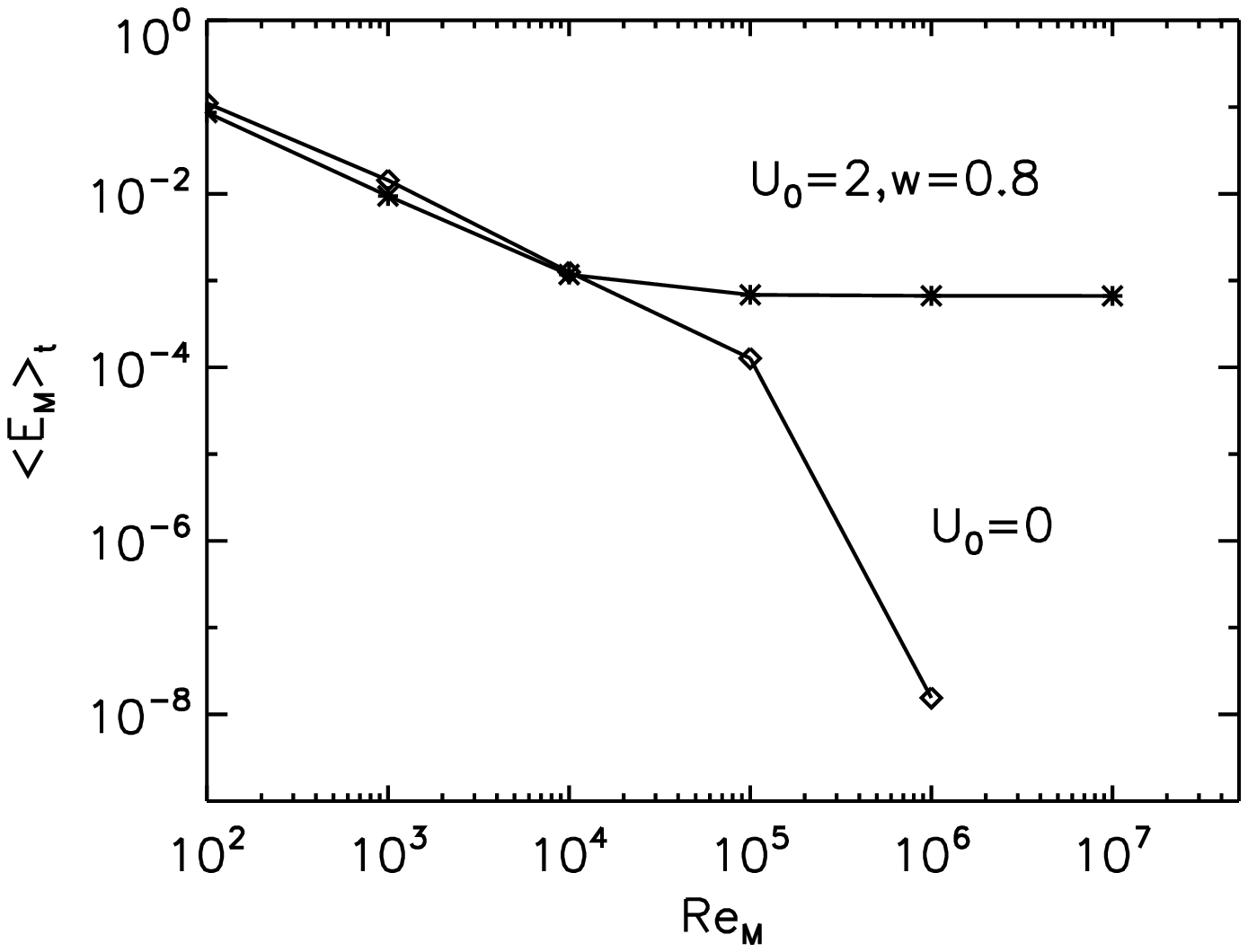}
\caption{The behaviour of the  time-averaged magnetic energy as a function
of magnetic Reynolds number $\Rm$, which shows the alleviation of
quenching, with wind speed $U_0=2$ and depth parameter $w=0.8$ (upper panel).
Also shown is the corresponding plot in the absence of a
wind, which clearly shows a catastrophic quenching.
The lower panel is for a $\alpha^2\Omega$ dynamo.}
\label{fig:alpha2}
\end{figure}

Finally we note that the alleviation of quenching as
described here is independent of some details of 
the  underlying dynamo model. In particular it does not
depend on whether we have an $\alpha^2$ dynamo
or an $\alpha^2\Omega$ dynamo. To check
this assertion explicitly we also solved the same problem
but with $\Ushear=0$ in Eq.~(\ref{eq:ubar}).
The results are shown in Fig.~\ref{fig:alpha2}
where we compare the alleviation of quenching
for the $\alpha^2$ dynamo (top panel) against the corresponding
$\alpha^2\Omega$ dynamo (bottom panel).
%between the two cases. 
%AB: this was forgotten to be removed
%%%%%%%%%%%%%%%%%%%%%%%%%%%%%%%%%%%%%%%%
%%%%%%%%%%%%%%%%%%%%%%%%%%%%%%%%%%%%%%%%

\section{Conclusions}
\label{Conclusions}
%--------------------
We have introduced two  observationally motivated effects that 
may help reduce the catastrophic quenching found in mean field 
dynamo models. An outward flow from the dynamo region (``wind'')
is found to be effective in allowing the quenching to saturate 
at finite values of the field strength. The wind
alone is, however, only effective when it penetrates quite deeply
into the convection zone.
These effects are modified to some extent by the presence of a meridional
circulation which has the ability to transport small scale helicity from
deep in the convection zone to near the surface, from where the wind can
more effectively remove it. However, the effects of circulation in our model
are not dramatic.
It is also true that the saturation fields in our model are rather
small compared to the equipartition field strength. This was also observed in
the model of \cite{SSSB06}; see also \cite{MS2010}. 
One possibility, that we have not explored, is that the neglected inhomogeneity of
the solar convection zone may be important.

It is interesting to try to estimate the various parameters of our model in
physical units and to compare them with
the solar values. We have taken the solar radius as the unit of
length ($7\times10^{10}$ cm).
The $\alpha$ effect can be taken to be a measure 
of the small scale velocity in the Sun, $\alpha \sim (1/3) |{\bm u}|$,
where ${\bm u} = {\bm U} - \Ubar$. 
The \cite{BT66} tables give estimates for small-scale
velocities in the convection zone of the Sun of between 
$4\times 10^3$--$2\times 10^5$ cm s$^{-1}$, 
in regions where convection is efficient. 
As we have considered
the convection zone to be homogeneous we consider 
$10^4$ cm s$^{-1}$ to be a reasonable estimate. 
Then, as $\alpha = 16$  in our units the unit of velocity is
$\sim 10^4/(3\times 16)$ cm s$^{-1} \sim 2 \times 10^3$ cm s$^{-1}$,
and the unit of time, obtained from length and velocity units
given above, is  $\sim 10^8$ s $\approx 10$ yrs.
Thus our characteristic cycle period, $T \approx 1$, 
corresponds to approximately 10 years.  
Then the maximum wind speed we have used ($U_0=10$) would correspond to
$2\times10^3$ cm s$^{-1}$. The speed of the meridional circulation at 
the surface in our units is $\vsurf = 0.47 $ for $\vamp=75$. 
Translated to physical units this becomes
$\vsurf \approx 1$ m s$^{-1}$,
which is of the same order of magnitude as the solar meridional velocity.
If in the estimates above we use the maximum and minimum values of the
small-scale velocity as given by the Baker \& Temesvary tables,
instead of the mean, the maximum surface speed of meridional circulation
will be between $0.4$ m s$^{-1}$  and $20$m s$^{-1}$ .
The speed of the solar wind that we have used is significantly smaller
than that of the actual solar wind, but on the other hand the real solar wind
is a highly fluctuating turbulent flow, whereas we have considered 
a constant outflow.

To summarise, we have presented a very simplified model, 
in order to explore some basic
ideas relevant to the solar dynamo. 
We cannot claim to have ``solved'' the quenching problem, but feel
we have identified, and to some extent quantified, mechanisms
of potential interest. We appreciate that there
are a number of desirable improvements, even in this MF formulation.
These include using a more realistic solar-like rotation law,
investigation and comparison of the effects of other fluxes of 
magnetic helicity \citep[e.g.][]{ZSRMLKK06}, the diffusive magnetic helicity
flux \citep{MCCTB2010}, the inclusion of compressibility in some form, but
most importantly perhaps, using a more realistic model for the solar wind
allowing for magnetic helicity loading via coronal mass ejections.
Notwithstanding these possible shortcomings, we do feel that our results
provide motivation for further investigations in the context of solar
and stellar dynamos.
Investigations using DNS \citep[e.g.][]{WB10} appear likely to be 
especially interesting, and we hope to pursue this approach. 

\begin{acknowledgements} 
%AB: added this
We thank an anonymous referee for suggesting several improvements
to the paper.
This work was supported by the the Leverhulme Trust,
the European Research Council under the AstroDyn Research Project 227952,
and the Swedish Research Council grant 621-2007-4064.
Computational resources were granted 
by QMUL HPC facilities purchased
under the SRIF initiative.
\end{acknowledgements}

%%----------------------------------------------

\end{document}